\documentclass[iop,revtex4]{emulateapj}
\usepackage{natbib}
\shorttitle{Astrometric Confirmation of 51 Eri b}
\shortauthors{De Rosa et al.}

\begin{document}
\submitted{Accepted November 4, 2015}
\title{Astrometric Confirmation and Preliminary Orbital Parameters\\ of the Young Exoplanet 51 Eridani \lowercase{b}\\ with the Gemini Planet Imager}

\author{
Robert J. De Rosa\altaffilmark{1},
Eric L. Nielsen\altaffilmark{2,3},
Sarah C. Blunt\altaffilmark{2,4}
James R. Graham\altaffilmark{1},
Quinn M. Konopacky\altaffilmark{5},\\
Christian Marois\altaffilmark{6,7},
Laurent Pueyo\altaffilmark{8},
Julien Rameau\altaffilmark{9},
Dominic M. Ryan\altaffilmark{1},
Jason J. Wang\altaffilmark{1},
Vanessa Bailey\altaffilmark{3},\\
Ashley Chontos\altaffilmark{3},
Daniel C. Fabrycky\altaffilmark{10},
Katherine B. Follette\altaffilmark{3},
Bruce Macintosh\altaffilmark{3},
Franck Marchis\altaffilmark{2},\\
S. Mark Ammons\altaffilmark{11},
Pauline Arriaga\altaffilmark{12},
Jeffrey K. Chilcote\altaffilmark{13},
Tara H. Cotten\altaffilmark{14}
Ren\'{e} Doyon\altaffilmark{9},\\
Gaspard Duch\^{e}ne\altaffilmark{1,15},
Thomas M. Esposito\altaffilmark{1},
Michael P. Fitzgerald\altaffilmark{12},
Benjamin Gerard\altaffilmark{7},\\
Stephen J. Goodsell\altaffilmark{16, 17},
Alexandra Z. Greenbaum\altaffilmark{18},
Pascale Hibon\altaffilmark{19},
Patrick Ingraham\altaffilmark{20},\\
Mara Johnson-Groh\altaffilmark{7},
Paul G. Kalas\altaffilmark{1},
David Lafreni\`{e}re\altaffilmark{9},
Jerome Maire\altaffilmark{13},
Stanimir Metchev\altaffilmark{21, 22},\\
Maxwell A. Millar-Blanchaer\altaffilmark{13},
Katie M. Morzinski\altaffilmark{23},
Rebecca Oppenheimer\altaffilmark{24},
Rahul I. Patel\altaffilmark{22},\\
Jennifer L. Patience\altaffilmark{25},
Marshall D. Perrin\altaffilmark{8},
Abhijith Rajan\altaffilmark{25},
Fredrik T. Rantakyr\"{o}\altaffilmark{19},\\
Jean-Baptiste Ruffio\altaffilmark{3},
Adam C. Schneider\altaffilmark{26},
Anand Sivaramakrishnan\altaffilmark{8},
Inseok Song\altaffilmark{14}, 
Debby Tran\altaffilmark{5},\\
Gautam Vasisht\altaffilmark{27},
Kimberly Ward-Duong\altaffilmark{25},
Schuyler G. Wolff\altaffilmark{18}}

\altaffiltext{1}{Astronomy Department, University of California, Berkeley, CA 94720, USA}
\altaffiltext{2}{SETI Institute, Carl Sagan Center, 189 Bernardo Avenue, Mountain View, CA 94043, USA}
\altaffiltext{3}{Kavli Institute for Particle Astrophysics and Cosmology, Stanford University, Stanford, CA 94305, USA}
\altaffiltext{4}{Department of Physics, Brown University, Providence, RI 02912, USA}
\altaffiltext{5}{Center for Astrophysics and Space Science, University of California San Diego, La Jolla, CA 92093, USA}
\altaffiltext{6}{National Research Council of Canada Herzberg, 5071 West Saanich Road, Victoria, BC V9E 2E7, Canada}
\altaffiltext{7}{Department of Physics and Astronomy, University of Victoria, 3800 Finnerty Road, Victoria, BC, V8P 5C2, Canada}
\altaffiltext{8}{Space Telescope Science Institute, 3700 San Martin Drive, Baltimore, MD 21218, USA}
\altaffiltext{9}{Institut de Recherche sur les Exoplan\`{e}tes, D\'{e}partment de Physique, Universit\'{e} de Montr\'{e}al, Montr\'{e}al QC H3C 3J7, Canada}
\altaffiltext{10}{Department of Astronomy and Astrophysics, University of Chicago, 5640 South Ellis Avenue, Chicago, IL 60637, USA}
\altaffiltext{11}{Lawrence Livermore National Laboratory, L-210, 7000 East Avenue, Livermore, CA 94550, USA}
\altaffiltext{12}{Department of Physics and Astronomy, University of California Los Angeles, 430 Portola Plaza, Los Angeles, CA 90095, USA}
\altaffiltext{13}{Dunlap Institute for Astronomy and Astrophysics, University of Toronto, Toronto, ON, M5S 3H4, Canada}
\altaffiltext{14}{Department of Physics and Astronomy, University of Georgia, Athens, GA 30602, USA}
\altaffiltext{15}{Universit\'{e} Grenoble Alpes / CNRS, Institut de Planétologie et d'Astrophysique de Grenoble, F-38000 Grenoble, France}
\altaffiltext{16}{Department of Physics, Durham University, Stockton Road, Durham, DH1 3LE, UK}
\altaffiltext{17}{Gemini Observatory, 670 N. A'ohoku Place, Hilo, HI 96720, USA}
\altaffiltext{18}{Department of Physics and Astronomy, Johns Hopkins University, Baltimore, MD 21218, USA}
\altaffiltext{19}{Gemini Observatory, Casilla 603, La Serena, Chile}
\altaffiltext{20}{Large Synoptic Survey Telescope, 950N Cherry Avenue, Tucson, AZ 85719, USA}
\altaffiltext{21}{Department of Physics and Astronomy, Centre for Planetary Science and Exploration, The University of Western Ontario, London, ON N6A 3K7, Canada}
\altaffiltext{22}{Department of Physics and Astronomy, Stony Brook University, 100 Nicolls Road, Stony Brook, NY 11790, USA}
\altaffiltext{23}{Steward Observatory, 933 N. Cherry Avenue, University of Arizona, Tucson, AZ 85721, USA}
\altaffiltext{24}{American Museum of Natural History, New York, NY 10024, USA}
\altaffiltext{25}{School of Earth and Space Exploration, Arizona State University, P.O. Box 871404, Tempe, AZ 85287, USA}
\altaffiltext{26}{Department of Physics and Astronomy, University of Toledo, 2801 W. Bancroft Street, Toledo, OH 43606, USA}
\altaffiltext{27}{Jet Propulsion Laboratory, California Inst. of Technology, 4800 Oak Grove Drive, Pasadena, CA 91009, USA}

\begin{abstract}
We present new Gemini Planet Imager observations of the young exoplanet 51~Eridani~b which provide further evidence that the companion is physically associated with 51~Eridani. Combining this new astrometric measurement with those reported in the literature, we significantly reduce the posterior probability that 51~Eridani~b is an unbound foreground or background T-dwarf in a chance alignment with 51~Eridani to $2\times10^{-7}$, an order of magnitude lower than previously reported. If 51~Eridani~b is indeed a bound object, then we have detected orbital motion of the planet between the discovery epoch and the latest epoch. By implementing a computationally efficient Monte Carlo technique, preliminary constraints are placed on the orbital parameters of the system. The current set of astrometric measurements suggest an orbital semimajor axis of $14^{+7}_{-3}$~AU, corresponding to a period of $41^{+35}_{-12}$~years (assuming a mass of $1.75$~$M_{\odot}$ for the central star), and an inclination of $138^{+15}_{-13}$~deg. The remaining orbital elements are only marginally constrained by the current measurements. These preliminary values suggest an orbit which does not share the same inclination as the orbit of the distant M-dwarf binary, GJ~3305, which is a wide physically bound companion to 51~Eridani.
\end{abstract}


\section{Introduction}
Monitoring the orbital motion of exoplanets, through direct imaging (e.g., \citealp{Chauvin:2012dk, Kalas:2013hp, Bonnefoy:2014bx, Nielsen:2014js}), or through indirect techniques such as radial velocity and transit measurements (e.g., \citealp{Cumming:2008hg,Howard:2012di,Marcy:2014hr,Moutou:2015fd}), can provide a wealth of information about their properties, the processes through which they form, and how they interact dynamically with other bodies in the system. Accurately determining the orbital parameters of exoplanets can constrain their masses and densities (e.g., \citealp{Charbonneau:2000fh}) and lead either to the discovery of additional planets in the system (e.g., \citealp{Nesvorny:2012gs}), or to the exclusion of additional planets within a range of periods by invoking dynamical stability arguments (e.g., \citealp{Correia:2005js}). Orbital parameters also provide insight as to how planetary companions dynamically interact with circumstellar material (e.g., \citealp{MillarBlanchaer:2015ha}). While the orbital periods are typically decades or longer for directly imaged planets, accurate astrometric monitoring of these systems can lead to preliminary constraints on their orbital parameters before a significant portion of the orbit is observed (e.g., \citealp{Pueyo:2015cx}).

51 Eridani (51 Eri) is a nearby ($29.43\pm0.29$~pc; \citealp{vanLeeuwen:2007dc}) member of the young ($24\pm3$~Myr; \citealp{Bell:2015gw}) $\beta$~Pictoris moving group \citep{Zuckerman:2001go}. Recently, \citet{Macintosh:2015ew} reported the discovery of a low-mass (2--10~$M_{\rm Jup}$) planet at a projected separation of $13.2\pm0.2$~AU based on observations with the Gemini Planet Imager (GPI; \citealp{Macintosh:2014js}). Based on $J$- and $H$-band spectroscopy, and $L^{\prime}$ photometry, 51~Eri~b was shown to have a spectrum with strong methane and water absorption, with a temperature of 600--750~K \citep{Macintosh:2015ew}. Due to the short baseline between discovery and follow-up, it was only possible to rule out a stationary background object from a 2003 non-detection; closer brown dwarf interlopers with non-zero proper motions could not be excluded. A statistical argument based on the space density of T-dwarfs \citep{Reyle:2010gq,Burningham:2013gt}, and the allowed range of distances of a foreground or background object based on the apparent magnitude of 51~Eri~b combined with the luminosity of T-dwarfs \citep{Kirkpatrick:2012ha}, was used to limit the possibility of an unassociated field brown dwarf to a probability of $2.4\times 10^{-6}$.

In addition to the newly resolved planetary companion, 51~Eri has an infrared excess indicative of a circumstellar debris disk \citep{Patel:2014dp,RiviereMarichalar:2014jm}. The debris disk has yet to be spatially resolved, so its geometry is unconstrained. At a projected separation of $1960$~AU lies GJ~3305---an M-dwarf binary with a semimajor axis of $9.80\pm0.15$~AU \citep{Montet:2015vr}---which is co-moving with 51~Eri, forming a bound hierarchical system \citep{Feigelson:2006ie}. While the inclination of the GJ~3305 binary is well constrained ($i=92\fdg1\pm0\fdg2$; \citealp{Montet:2015vr}), the period of the wide 51~Eri--GJ~3305 binary ($\sim10^4$~years; \citealp{Feigelson:2006ie}) precludes any such estimation of its inclination.

\section{Observations and Data Reduction}
\begin{figure}
\center
\includegraphics[width=0.5\textwidth]{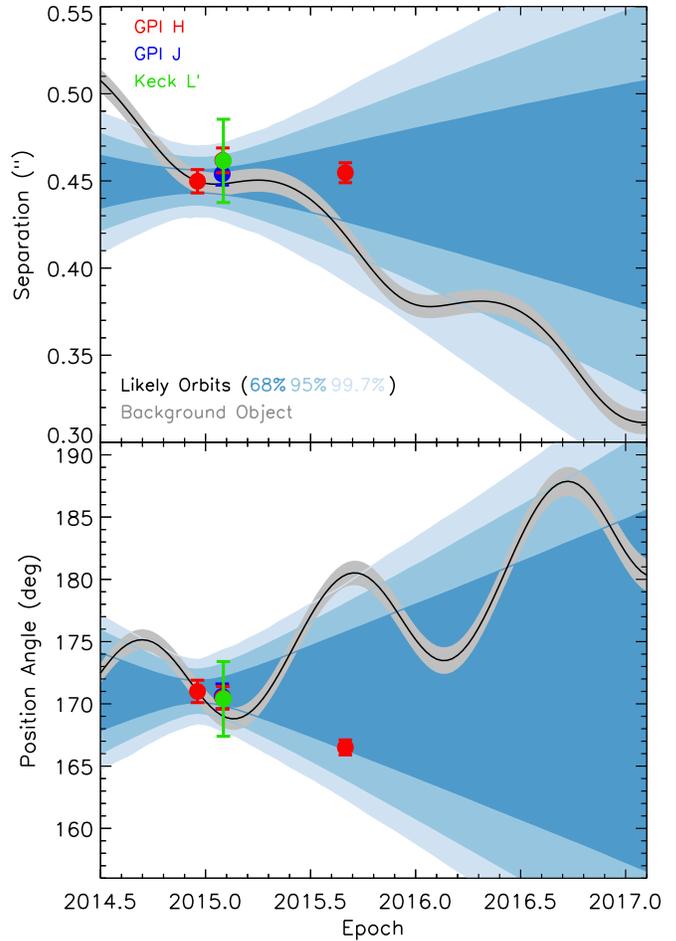}
\caption{Astrometry of 51~Eri~b from 2014 December to 2015 September (filled red, blue, and green circles). The non-moving background object hypothesis (light gray track), computed from the {\it Hipparcos}-measured parallax and proper motion of 51~Eri, is robustly rejected using these data alone. The measured displacement is well within the range of orbital motion expected for bound planetary-mass companions (blue envelopes). Likely orbital tracks were generated using the Monte Carlo method described in Section~\ref{orbit_sec} to produce $10^4$ orbits fit to the first epoch, with the plotted ranges encompassing 68\%, 95\%, and 99.7\% of the orbits.}
\label{fig:cpm_bkg}
\end{figure}
51~Eri was initially observed with GPI at Gemini South as a part of the GPI Exoplanet Survey (GPIES) on 2014 December 18 UT (GS-2014B-Q-500). A faint companion candidate was identified, and subsequent observations demonstrated that its spectral energy distribution was consistent with that of a low-temperature, low-surface gravity giant planet \citep{Macintosh:2015ew}. In total, 51~Eri~b has been successfully observed four times since discovery in 2014, three times with GPI (GS-2014B-Q-501, GS-2015A-Q-501), and once with NIRC2 at the W.~M.~Keck~2 telescope using the facility adaptive optics system \citep{Wizinowich:2000hl}. A summary of these observations is given in Table \ref{tab:obs}. For each GPI epoch, the observing strategy was the same. The target was acquired before transit to maximize field rotation \citep{Marois:2006df}, and the observations were taken using the spectral coronagraphic mode, with either the $J$- or $H$-band filters. In addition to these successful observations, 51~Eri was also observed with GPI in $J$-band on 2015 January 29, where 51~Eri~b was not recovered due to poor image quality, and in $H$-band on 2015 August 30 and 31 where, although 51~Eri~b was recovered, the signal-to-noise ratio was significantly worse.

\setcounter{footnote}{0}
The GPI observations obtained on 2015 September 1 were reduced using the GPI Data Reduction Pipeline (DRP; \citealp{Perrin:2014jh})\footnote{\tt http://docs.planetimager.org/pipeline/}. The dark current was subtracted and bad pixels were identified and fixed. The shift in the position of the micro spectra on the detector due to mechanical flexure was measured by comparing reference argon arcs taken monthly, to arcs obtained after target acquisition \citep{Wolff:2014cn}. The micro spectra were extracted, converting the 2D image into a 3D $(x, y, \lambda)$ datacube. These were then divided by a flat field to correct for lenslet throughput, and were interpolated along the wavelength axis to a common wavelength vector across the bandpass. Finally, the optical distortion was corrected for using measurements obtained with a pinhole mask \citep{Konopacky:2014hf}.

To minimize potential biases between the astrometry presented in \citet{Macintosh:2015ew} and the new measurements presented here, we used Pipelines 1 and 3 from \citet{Macintosh:2015ew} to both perform the point-spread function (PSF) subtraction, and to extract the astrometry of 51~Eri~b. Pipeline 2 was switched to a Python implementation \citep{Wang:2015th} of the Karhunen--Lo\`{e}ve Image Projection algorithm \citep{Soummer:2012ig}, and uses a forward-modeled PSF to perform a Markov Chain Monte Carlo (MCMC) analysis to determine the posterior distributions of the separation and position angle of 51~Eri~b.

The plate scale and position angle of GPI have been monitored by continually observing a set of astrometric calibrators with well-determined orbital solutions or contemporaneous NIRC2 measurements, which has an accurate astrometric solution \citep{Yelda:2010ig}. In addition to observations listed in \citet{Konopacky:2014hf}, we have observed the $\theta^1$~Ori~B quadruple system an additional four times, the HD~157516 binary twice, and the HIP~80628 binary once. These observations were reduced as above, and PSF subtraction was not required. The pixel positions of each component were measured as in \citet{Konopacky:2014hf}.

Combining these measurements results in a plate scale of $14.166\pm0.007$~mas~lenslet$^{-1}$, and a position angle offset of $-1\fdg10\pm0\fdg13$. The position angle offset is defined as the angle between the lenslet $y$-axis, and the elevation axis of the telescope, measured east from north. Since version 1.2, the DRP takes into account an offset of $-1^{\circ}$ while processing the data, and as such the difference between the true position angle and the measured position angle in a DRP-reduced image is $\theta_{\rm true} - \theta_{\rm measured} = -0\fdg10\pm0\fdg13$. There is no evidence of variations of either the plate scale or position angle offset between observing runs within the measurement uncertainties, so a single value is adopted for all epochs. The revised astrometric calibration was used to recalculate the astrometry from \citet{Macintosh:2015ew}, which are consistent with the previous values, and are shown in Table~\ref{tab:obs}.

\section{Common Proper Motion Confirmation}
\begin{figure}
\center
\includegraphics[width=0.5\textwidth,trim=1.5cm 0 0 0]{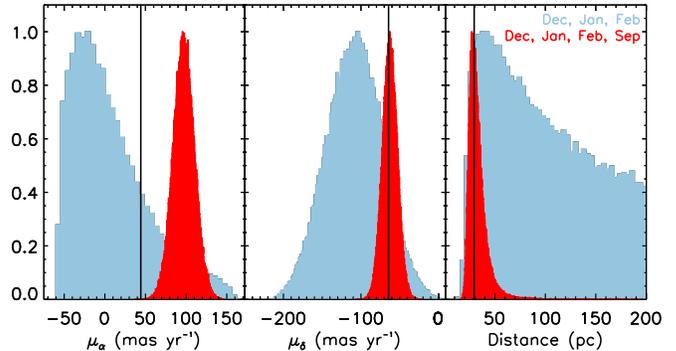}
\caption{Normalized posterior distributions of the proper motion and distance of 51~Eri~b assuming it is an unbound brown dwarf, derived from the epochs in \citet{Macintosh:2015ew} using the updated astrometric calibration and the 2003 non-detection (blue histogram). The addition of the 2015 September measurement significantly improves the constraint on each parameter (red histogram). The proper motion and distance to 51~Eri are denoted by the vertical lines. The distance posterior distribution is strongly peaked at the distance of 51~Eri. Using this constraint, the posterior probability that 51~Eri~b is an unbound field dwarf is calculated as $2\times 10^{-7}$.}
\label{fig:plx_pm_fit}
\end{figure}
Typically, confirmation of common proper motion is achieved by comparing the motion of a candidate with respect to a background track for a stationary background object with negligible parallax, so that the only relevant movement is the parallax and proper motion of the primary star (e.g., \citealp{Nielsen:2013jy}). Such an analysis for 51~Eri~b is shown in Figure~\ref{fig:cpm_bkg}, with the astrometry diverging from the stationary background object track ($2.8$-$\sigma$ in $\rho$, $8.8$-$\sigma$ in $\theta$). The spectrum of 51~Eri~b excluded the distant background star hypothesis, leaving two plausible fits to the spectrum: an unbound field brown dwarf, or a bound planet, as described by \citet{Macintosh:2015ew}, who found a probability of $2.4 \times 10^{-6}$ that the object was an unbound field brown dwarf. By including the latest epoch, this probability can be further reduced by solving for the allowable parallax and proper motion of an unbound object, and reducing the volume in which an unbound T-dwarf could exist.

Using the measured parallax and proper motion of 51~Eri \citep{vanLeeuwen:2007dc} the relative astrometry of 51~Eri~b was converted into absolute astrometry. A Metropolis--Hastings MCMC technique (e.g., \citealt{Ford:2006ej}) was then used to fit the proper motion and parallax of 51~Eri~b to both the absolute astrometry, and the non-detection from 2003 discussed in \citet{Macintosh:2015ew}. A uniform prior in the two proper motion directions was used, and a $p(d)\propto d^2$ prior for the distance, with a maximum distance of 200~pc. Good convergence was achieved in the MCMC chains, with a Gelman--Rubin statistic $< 1.00007$. The posteriors from the MCMC fit are shown in Figure~\ref{fig:plx_pm_fit}.  The median distance from the posteriors is 31~pc, with a 68\% confidence interval between 25 and 39~pc, consistent with the distance to 51~Eri ($29.4\pm0.3$~pc). The offset in proper motion with respect to the star is consistent with orbital motion occurring over the span of the observations.

\begin{figure*}
\center
\includegraphics[width=\textwidth,trim={0 8.5cm 0 0}]{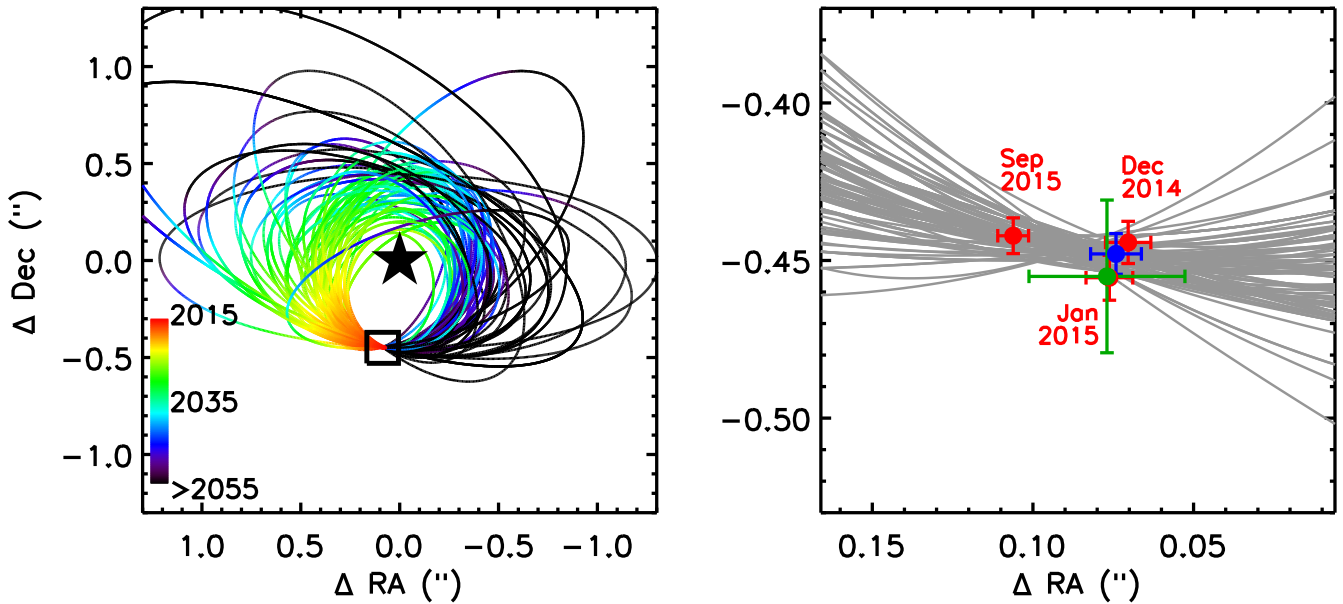}
\caption{(left): One hundred randomly selected orbits for 51~Eri~b from the Monte Carlo technique described in Section~\ref{orbit_sec} consistent with the astrometry. The color corresponds to the epoch of a given location along each orbital track, epochs later than 2055 are plotted in black. The position of 51~Eri is indicated by the black star. The region plotted in the right hand side is indicated by the black square (right): As the left panel, but focusing on the available astrometry. The color on the orbital tracks has been removed for clarity. The color of the symbols are as in Figure 1, and the multiple GPI $H$-band epochs are labelled.}
\label{fig:firework}
\end{figure*}
The calculation of the unbound brown dwarf probability from \citet{Macintosh:2015ew} is updated using this new distance constraint. The previous calculation found the product of the number density of T-dwarfs and the volume of a cone with the angular width of the GPI detector out to the largest distance a T-dwarf could be seen in the reduced GPI image of a given target, repeated for each of the 44 stars observed as a part of GPIES at the time of the discovery.  There was a further correction factor based on the proper motion constraints from the 2003 non-detection, as only 34\% of the expected distribution of field brown dwarfs had the motion required to place the object behind the star in 2003.  This correction is not used here as the revised proper motion estimate for the unbound brown dwarf scenario is 3-$\sigma$ discrepant from the proper motion of the star due to orbital motion (Figure~\ref{fig:plx_pm_fit}). Instead, the volume calculation of the cones was convolved with the distance posterior distribution, lowering the posterior probability that 51~Eri~b is an unbound brown dwarf to $2\times 10^{-7}$.

\section{Orbital Elements of 51 Eri \lowercase{b}}\label{orbit_sec}
\begin{figure*}
\center
\includegraphics[width=\textwidth]{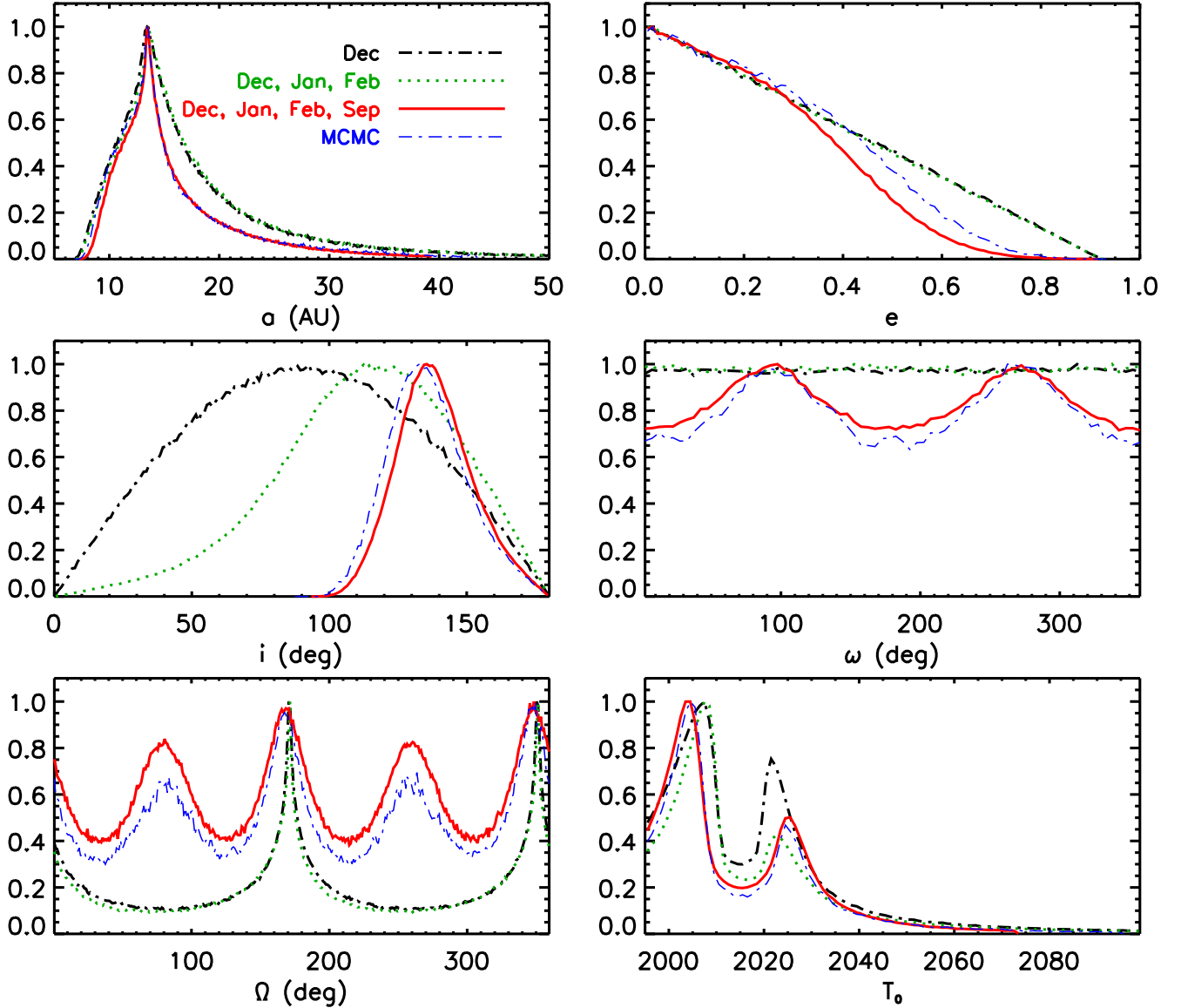}
\caption{Normalized distributions of accepted orbital parameters consistent with the astrometry from the Monte Carlo analysis described in Section~\ref{orbit_sec}. The analysis was performed using the discovery epoch (black dashed histogram), using all epochs presented in \citet{Macintosh:2015ew} (green dotted histogram), and using all epochs (red solid histogram). As more astrometry is added, the inclination angle becomes significantly more constrained. The normalized posterior distributions from the MCMC analysis are also shown for comparison (blue dotted--dashed histogram).}
\label{fig:orbit_params}
\end{figure*}
\begin{deluxetable*}{ccccccccccc}
\tabletypesize{\scriptsize}
\tablecaption{Observations and Orbital Parameters of 51~Eridani~\lowercase{b}}
\tablewidth{0pt}
\tablehead{
\colhead{UT Date} & \colhead{MJD} & \colhead{Instrument} & \colhead{Filter} & $t_{\rm int}$ & $N_{\rm coadd}$ & $N_{\rm exp}$  &\colhead{Plate Scale} & \colhead{Position Angle} & \colhead{$\rho$} & \colhead{$\theta$}\\
& & & & (s) & & &  (mas~px$^{-1}$)$^{\rm a}$ & Offset (deg) & (mas) & (deg)
}
\startdata
2014 Dec 18 & 57009.1292 & GPI & $H$ & 59.6 & 1 & 38 & $14.166\pm0.007$ & $-1.10\pm0.13$ & $449.8\pm6.7$& $171.0\pm0.9$\\
2015 Jan 30 & 57052.0572 & GPI & $J$ & 59.6 & 1 & 70 & $14.166\pm0.007$ & $-1.10\pm0.13$ & $454.0\pm6.4$ & $170.6\pm1.0$\\
2015 Jan 31 & 57052.9753 & GPI & $H$ & 59.6 & 1 &64 & $14.166\pm0.007$ & $-1.10\pm0.13$ & $461.8\pm7.1$ & $170.5\pm0.9$\\
2015 Feb 01 & 57054.0364 & NIRC2 & $L^\prime$ & 0.9 & 60 & 62 & $9.952\pm0.002^{\rm b}$ & $0.252\pm0.009^{\rm b}$ & $461.5\pm23.9$ & $170.4\pm3.0$\\
2015 Sep 01 & 57266.4052 & GPI & $H$ & 59.6 & 1 & 93 & $14.166\pm0.007$ & $-1.10\pm0.13$ & $454.7\pm5.7$ & $166.5\pm0.6$\\
\\
\hline
\hline
\multicolumn{11}{c}{\footnotesize Preliminary Orbital Parameters of 51~Eridani~b}\\
\multicolumn{2}{c}{Parameter} & Unit & \multicolumn{2}{c}{$\chi^2_{\rm min}$} & \multicolumn{2}{c}{Median} & Lower & Upper&\\
\hline
\multicolumn{2}{c}{Semimajor axis ($a$)} & AU & \multicolumn{2}{c}{18.61} & \multicolumn{2}{c}{14} & 11 & 21\\
\multicolumn{2}{c}{Eccentricity ($e$)} & - & \multicolumn{2}{c}{0.2804} & \multicolumn{2}{c}{0.21} & 0.06 & 0.40\\
\multicolumn{2}{c}{Inclination ($i$)} & deg & \multicolumn{2}{c}{126.7} & \multicolumn{2}{c}{138} & 125 & 153\\
\multicolumn{2}{c}{Argument of periastron ($\omega$)} & deg & \multicolumn{2}{c}{4.233} & \multicolumn{2}{c}{90} & 32 & 148\\
\multicolumn{2}{c}{Position angle of nodes ($\Omega$)} & deg & \multicolumn{2}{c}{166.6} & \multicolumn{2}{c}{75} & 47 & 100\\
\multicolumn{2}{c}{Epoch of periastron ($T_0$)} & - & \multicolumn{2}{c}{2016.05} & \multicolumn{2}{c}{2014.79} & 2003.31 & 2026.11\\
\multicolumn{2}{c}{Period ($P$)} & year & \multicolumn{2}{c}{60.68} & \multicolumn{2}{c}{41} & 29 & 76\\
\enddata
\tablenotetext{a}{\citet{Yelda:2010ig}}
\tablenotetext{b}{In reduced GPI datacubes, one pixel is equivalent to one lenslet}
\label{tab:obs}
\end{deluxetable*}
MCMC orbit fitting is slow to converge for sparsely sampled astrometry or short orbital arcs, and so we implemented a more computationally efficient Monte Carlo method to generate plausible distributions of orbital parameters based on astrometry covering only a small fraction of an orbital period. For four of the orbital parameters, a large number of samples were drawn from appropriate probability distributions: uniform for argument of periastron ($\omega$) and epoch of periastron passage ($T_0$), uniform in $\cos(i)$ for inclination angle~($i$), and eccentricity ($e$) following the linear fit to radial velocity planets of \citet{Nielsen:2008kk}. Initial orbits were generated using these four parameter distributions and fixed values of semimajor axis ($a$) and position angle of nodes ($\Omega$), the values of $a$ was scaled and the value of $\Omega$ rotated to reproduce the astrometry at the first epoch. Period ($P$) was not fit and was calculated assuming a stellar mass of 1.75 $M_{\odot}$ \citep{Simon:2011ix}. Astrometric errors were incorporated into the generated parameters by adding random offsets to the separation and position angle of the first epoch before each orbit is shifted and rotated.  These uncertainties were randomly drawn from Gaussian distributions equivalent to the first epoch astrometric uncertainties.

With multiple epochs, we proceeded to iteratively reject sets of orbital parameters that do not match later measurements, with acceptance probability given by a two-dimensional Gaussian with the astrometric errors as the standard deviations.  Generated orbits that were closest to the observed separation and position angle at the corresponding observational epoch were more likely to be accepted. We then obtained distributions of fitting orbital parameters given the input astrometry. Varying the epoch chosen to initialize the procedure had little effect on the distribution of accepted orbital parameters.

We validated this method by generating $10^3$ orbits with one fixed orbital parameter and randomly sampling the other parameters, and then creating five epochs from one year of simulated astrometry for each orbit, each with an observational uncertainty.  The spacing of the epochs and astrometric errors were chosen to be the same as the measurements of 51~Eri~b.  We then applied our method to the artificial astrometry and examined the returned distributions of orbital parameters.  Of the simulated orbits with semimajor axis of 13~AU, the median was 13.3~AU, and in 74\% of orbits 13~AU was within the 68\% confidence interval of each individual orbital fit for semimajor axis.  This suggests the generated distributions are reasonable representations of the posterior probability distributions, and comparison with MCMC (described below) further supports this.  This procedure will be described further in Blunt et al. (2015, in preparation).

This technique was applied to the astrometry of 51~Eri~b. One hundred orbits from the fit are shown in Figure~\ref{fig:firework}, and the distributions of accepted orbital parameters are shown in Figure~\ref{fig:orbit_params}. While the marginal distribution for inclination using the discovery epoch follows the shape of the prior, the distribution after incorporating data from 2015 peaks at $\sim138^{\circ}$, showing the new observations have provided significant orbital constraints.  More moderate changes are seen in eccentricity and semimajor axis, with less eccentric orbits, and orbits closer to 14 AU, becoming more favored with increasing number of measurements. The orbital elements corresponding to the minimum $\chi^2$, representing the best-fitting orbit generated, are given in Table~\ref{tab:obs}, along with the median and limits corresponding to where 68\% of the Monte Carlo generated orbits were found.  In order to compute the median and 68\% intervals, $\omega$ was wrapped to be within $0^{\circ}$--$180^{\circ}$, $\Omega$ to within $30^{\circ}$--$120^{\circ}$, and $T_0$ to within 1995--1995$+P$. 

The shapes of the distributions of accepted orbital parameters were confirmed using an affine invariant MCMC ensemble sampler \citep{ForemanMackey:2013io}\footnote{http://dan.iel.fm/emcee}, with the same prior on each parameter. The shapes of the posterior distributions are similar (Figure~\ref{fig:orbit_params}); the median and 1-$\sigma$ confidence intervals for $a$ and $i$ were found to be $14.1^{+8.2}_{-3.0}$~AU and $135^{+15}_{-13}$~deg, consistent with the values in Table~\ref{tab:obs}, but taking two orders of magnitude longer to compute. In addition to these Monte Carlo techniques, the astrometry of 51~Eri~b was applied to the technique for constraining orbital parameters over short orbital arcs presented in \citet{Pearce:2015je}. The angle between the projected separation and velocity vectors was calculated as $\phi=84.5^{+14.2}_{-14.3}$~deg, and a value for their dimensionless parameter $B$ of $0.23^{+0.12}_{-0.09}$. Comparing these to the minimum inclination and eccentricity contours of \citet{Pearce:2015je}, the eccentricity of 51~Eri~b is unconstrained, and the inclination is restricted to $i < 77^{+5}_{-7}$~deg (or $i > 103^{+7}_{-5}$~deg). While consistent with the values in Table~\ref{tab:obs}, these limits are significantly less constraining.

\section{Conclusions}
We have presented a new astrometric measurement of the young exoplanet 51~Eridani~b, which provides further evidence supporting the bound companion hypothesis. Using this new measurement, we observe significant deviation ($2.8$-$\sigma$ in $\rho$, $8.8$-$\sigma$ in $\theta$) from the stationary background object track, with a reduced $\chi^2_{\rm bkg}$ of $11.19$ \citep{Nielsen:2012jk}. We compute a probability that it is an interloping field brown dwarf of $2\times10^{-7}$, decreased by a factor of twelve relative to \citet{Macintosh:2015ew}. By implementing a computationally efficient Monte Carlo method to sample probable orbits, we place the first constraints on the orbital parameters of 51~Eri~b. Based on the present astrometry, the median of the semimajor axis distribution is $14$~AU, corresponding to a period of $41$~years, assuming a mass of 1.75~$M_{\odot}$ for 51~Eri. While the additional astrometric epoch did not significantly change the distribution of accepted semimajor axes and periods, the range of allowed inclinations was significantly constrained, with a median value of $138^{\circ}$. The eccentricity of the orbit remains unconstrained, with circular orbits only marginally preferred relative to the prior distribution. 

Based on these preliminary constraints, the orbit of 51~Eri~b does not appear to be co-planar with the orbit of the M-dwarf binary GJ~3305 ($i=92\fdg1\pm0\fdg2$; \citealp{Montet:2015vr}). The inclination of the orbital plane of the outer binary relative to 51~Eri is not known. The large separation between 51~Eri and GJ~3305, and the young age of the system, would suggest that secular Lidov--Kozai oscillations would not have had sufficient time to significantly alter the semimajor axis of the planet \citep{Montet:2015vr}, although moderate changes in the inclination and eccentricity are not excluded \citep{Fabrycky:2007jh}. Given the long timescale of these oscillations (200~Myr for a perturber on a circular orbit, \citealp{Montet:2015vr}), this effect could be completely suppressed by quicker secular precession due to, e.g., a relatively low-mass, undetected planet \citep{Wu:2003kl}. Alternatively, the planet may remain on its current orbit, despite the presence of the wide binary \citep{Holman:1999cu}. Continued astrometric monitoring of 51~Eri~b over the next few years should be sufficient to detect curvature in the orbit, further constraining the semimajor axis and inclination of the orbit, and placing the first constraints on the eccentricity. Absolute astrometric measurements of 51~Eri with {\it GAIA} (e.g., \citealp{Perryman:2014jr}), in conjunction with monitoring of the relative astrometry of 51~Eri~b, will enable a direct measurement of the mass of the planet. Combined with the well-constrained age of 51~Eri~b, such a determination would provide insight into the evolutionary history of low-mass directly imaged extrasolar planets, and help distinguish between a hot-start or core accretion formation process for this planet.

\acknowledgments

Based on observations obtained at the Gemini Observatory, which is operated by the Association of Universities for Research in Astronomy, Inc., under a cooperative agreement with the National Science Foundation (NSF) on behalf of the Gemini partnership: the NSF (United States), the National Research Council (Canada), CONICYT (Chile), the Australian Research Council (Australia), Minist\'{e}rio da Ci\^{e}ncia, Tecnologia e Inova\c{c}\~{a}o (Brazil) and Ministerio de Ciencia, Tecnolog\'{i}a e Innovaci\'{o}n Productiva (Argentina). This research has made use of the SIMBAD database, operated at CDS, Strasbourg, France. Supported by NSF grants AST-0909188 and AST-1313718 (R.J.D.R., J.R.G., J.J.W., T.M.E., P.G.K.), AST-1411868 (B.M., K.F., J.L.P., A.R., K.W.D.), AST-141378 (P.A., G.D., M.P.F.), NNX11AF74G (A.Z.G., A.S.), and DGE-1232825 (A.Z.G.). Supported by NASA grants NNX15AD95G/NEXSS and NNX11AD21G (R.J.D.R., J.R.G., J.J.W., T.M.E., P.G.K.), and NNX14AJ80G (E.L.N., S.C.B., B.M., F.M., M.P.). J.R., R.D. and D.L. acknowledge support from the Fonds de Recherche du Qu\'{e}bec. Portions of this work were performed under the auspices of the U.S. Department of Energy by Lawrence Livermore National Laboratory under Contract DE-AC52-07NA27344 (S.M.A.). B.G. and M.J.G. acknowledge support from the National Sciences and Engineering Research Council of Canada. G.V. acknowledges a JPL Research and Technology Grant for improvements to the GPI CAL system.

{\it Facility:} \facility{Gemini:South (GPI)}.


\begin{thebibliography}{}
\expandafter\ifx\csname natexlab\endcsname\relax\def\natexlab#1{#1}\fi

\bibitem[{Bell {et~al.}(2015)Bell, Mamajek, \& Naylor}]{Bell:2015gw}
Bell, C. P.~M., Mamajek, E.~E., \& Naylor, T. 2015, MNRAS, 454, 593

\bibitem[{Bonnefoy {et~al.}(2014)Bonnefoy, Marleau, Galicher, Beust, Lagrange,
  Baudino, Chauvin, Borgniet, Meunier, Rameau, Boccaletti, Cumming, Helling,
  Homeier, Allard, \& Delorme}]{Bonnefoy:2014bx}
Bonnefoy, M., Marleau, G.~D., Galicher, R., {et~al.} 2014, A{\&}A, 567, L9

\bibitem[{Burningham {et~al.}(2013)Burningham, Cardoso, Smith, Leggett, Smart,
  Mann, Dhital, Lucas, Tinney, Pinfield, Zhang, Morley, Saumon, Aller,
  Littlefair, Homeier, Lodieu, Deacon, Marley, van Spaandonk, Baker, Allard,
  Andrei, Canty, Clarke, Day-Jones, Dupuy, Fortney, Gomes, Ishii, Jones, Liu,
  Magazz{\'u}, Marocco, Murray, Rojas-Ayala, \& Tamura}]{Burningham:2013gt}
Burningham, B., Cardoso, C.~V., Smith, L., {et~al.} 2013, MNRAS, 433, 457

\bibitem[{Charbonneau {et~al.}(2000)Charbonneau, Brown, Latham, \&
  Mayor}]{Charbonneau:2000fh}
Charbonneau, D., Brown, T.~M., Latham, D.~W., \& Mayor, M. 2000, ApJL, 529, L45

\bibitem[{Chauvin {et~al.}(2012)Chauvin, Lagrange, Beust, Bonnefoy, Boccaletti,
  Apai, Allard, Ehrenreich, Girard, Mouillet, \& Rouan}]{Chauvin:2012dk}
Chauvin, G., Lagrange, A.~M., Beust, H., {et~al.} 2012, A{\&}A, 542, 41

\bibitem[{Correia {et~al.}(2005)Correia, Udry, Mayor, Laskar, Naef, Pepe,
  Queloz, \& Santos}]{Correia:2005js}
Correia, A. C.~M., Udry, S., Mayor, M., {et~al.} 2005, A{\&}A, 440, 751

\bibitem[{Cumming {et~al.}(2008)Cumming, Butler, Marcy, Vogt, Wright, \&
  Fischer}]{Cumming:2008hg}
Cumming, A., Butler, R.~P., Marcy, G.~W., {et~al.} 2008, PASP, 120, 531

\bibitem[{Fabrycky \& Tremaine(2007)}]{Fabrycky:2007jh}
Fabrycky, D., \& Tremaine, S. 2007, ApJ, 669, 1298

\bibitem[{Feigelson {et~al.}(2006)Feigelson, Lawson, Stark, Townsley, \&
  Garmire}]{Feigelson:2006ie}
Feigelson, E.~D., Lawson, W.~A., Stark, M., Townsley, L., \& Garmire, G.~P.
  2006, AJ, 131, 1730

\bibitem[{Ford(2006)}]{Ford:2006ej}
Ford, E.~B. 2006, ApJ, 642, 505

\bibitem[{Foreman-Mackey {et~al.}(2013)Foreman-Mackey, Hogg, Lang, \&
  Goodman}]{ForemanMackey:2013io}
Foreman-Mackey, D., Hogg, D.~W., Lang, D., \& Goodman, J. 2013, PASP, 125, 306

\bibitem[{Holman \& Wiegert(1999)}]{Holman:1999cu}
Holman, M.~J., \& Wiegert, P.~A. 1999, AJ, 117, 621

\bibitem[{Howard {et~al.}(2012)Howard, Marcy, Bryson, Jenkins, Rowe, Batalha,
  Borucki, Koch, Dunham, Gautier, Van~Cleve, Cochran, Latham, Lissauer, Torres,
  Brown, Gilliland, Buchhave, Caldwell, Christensen-Dalsgaard, Ciardi, Fressin,
  Haas, Howell, Kjeldsen, Seager, Rogers, Sasselov, Steffen, Basri,
  Charbonneau, Christiansen, Clarke, Dupree, Fabrycky, Fischer, Ford, Fortney,
  Tarter, Girouard, Holman, Johnson, Klaus, Machalek, Moorhead, Morehead,
  Ragozzine, Tenenbaum, Twicken, Quinn, Isaacson, Shporer, Lucas, Walkowicz,
  Welsh, Boss, Devore, Gould, Smith, Morris, Prsa, Morton, Still, Thompson,
  Mullally, Endl, \& MacQueen}]{Howard:2012di}
Howard, A.~W., Marcy, G.~W., Bryson, S.~T., {et~al.} 2012, ApJS, 201, 15

\bibitem[{Kalas {et~al.}(2013)Kalas, Graham, Fitzgerald, \&
  Clampin}]{Kalas:2013hp}
Kalas, P., Graham, J.~R., Fitzgerald, M.~P., \& Clampin, M. 2013, ApJ, 775, 56

\bibitem[{Kirkpatrick {et~al.}(2012)Kirkpatrick, Gelino, Cushing, Mace,
  Griffith, Skrutskie, Marsh, Wright, Eisenhardt, McLean, Mainzer, Burgasser,
  Tinney, Parker, \& Salter}]{Kirkpatrick:2012ha}
Kirkpatrick, J.~D., Gelino, C.~R., Cushing, M.~C., {et~al.} 2012, ApJ, 753, 156

\bibitem[{Konopacky {et~al.}(2014)Konopacky, Thomas, Macintosh, Dillon,
  Sadakuni, Maire, Fitzgerald, Hinkley, Kalas, Esposito, Marois, Ingraham,
  Marchis, Perrin, Graham, Wang, De~Rosa, Morzinski, Pueyo, Chilcote, Larkin,
  Fabrycky, Goodsell, Oppenheimer, Patience, Saddlemyer, \&
  Sivaramakrishnan}]{Konopacky:2014hf}
Konopacky, Q.~M., Thomas, S.~J., Macintosh, B.~A., {et~al.} 2014,
  Proc.{\textasciitilde}SPIE, 9147, 84

\bibitem[{Macintosh {et~al.}(2014)Macintosh, Graham, Ingraham, Konopacky,
  Marois, Perrin, Poyneer, Bauman, Barman, Burrows, Cardwell, Chilcote,
  De~Rosa, Dillon, Doyon, Dunn, Erikson, Fitzgerald, Gavel, Goodsell, Hartung,
  Hibon, Kalas, Larkin, Maire, Marchis, Marley, McBride, Millar-Blanchaer,
  Morzinski, Norton, Oppenheimer, Palmer, Patience, Pueyo, Rantakyro, Sadakuni,
  Saddlemyer, Savransky, Serio, Soummer, Sivaramakrishnan, Song, Thomas,
  Wallace, Wiktorowicz, \& Wolff}]{Macintosh:2014js}
Macintosh, B., Graham, J.~R., Ingraham, P., {et~al.} 2014, PNAS, 111, 12661

\bibitem[{Macintosh {et~al.}(2015)Macintosh, Graham, Barman, De~Rosa,
  Konopacky, Marley, Marois, Nielsen, Pueyo, Rajan, Rameau, Saumon, Wang,
  Patience, Ammons, Arriaga, Artigau, Beckwith, Brewster, Bruzzone, Bulger,
  Burningham, Burrows, Chen, Chiang, Chilcote, Dawson, Dong, Doyon, Draper,
  Duch{\^e}ne, Esposito, Fabrycky, Fitzgerald, Follette, Fortney, Gerard,
  Goodsell, Greenbaum, Hibon, Hinkley, Cotten, Hung, Ingraham, Johnson-Groh,
  Kalas, Lafreni{\`e}re, Larkin, Lee, Line, Long, Maire, Marchis, Matthews,
  Max, Metchev, Millar-Blanchaer, Mittal, Morley, Morzinski, Murray-Clay,
  Oppenheimer, Palmer, Patel, Perrin, Poyneer, Rafikov, Rantakyr{\"o}, Rice,
  Rojo, Rudy, Ruffio, Ruiz, Sadakuni, Saddlemyer, Salama, Savransky, Schneider,
  Sivaramakrishnan, Song, Soummer, Thomas, Vasisht, Wallace, Ward-Duong,
  Wiktorowicz, Wolff, \& Zuckerman}]{Macintosh:2015ew}
Macintosh, B., Graham, J.~R., Barman, T., {et~al.} 2015, Science, 350, 64

\bibitem[{Marcy {et~al.}(2014)Marcy, Isaacson, Howard, Rowe, Jenkins, Bryson,
  Latham, Howell, Gautier~III, Batalha, Rogers, Ciardi, Fischer, Gilliland,
  Kjeldsen, Christensen-Dalsgaard, Huber, Chaplin, Basu, Buchhave, Quinn,
  Borucki, Koch, Hunter, Caldwell, Van~Cleve, Kolbl, Weiss, Petigura, Seager,
  Morton, Johnson, Ballard, Burke, Cochran, Endl, MacQueen, Everett, Lissauer,
  Ford, Torres, Fressin, Brown, Steffen, Charbonneau, Basri, Sasselov, Winn,
  Sanchis-Ojeda, Christiansen, Adams, Henze, Dupree, Fabrycky, Fortney, Tarter,
  Holman, Tenenbaum, Shporer, Lucas, Welsh, Orosz, Bedding, Campante, Davies,
  Elsworth, Handberg, Hekker, Karoff, Kawaler, Lund, Lundkvist, Metcalfe,
  Miglio, Aguirre, Stello, White, Boss, Devore, Gould, Prsa, Agol, Barclay,
  Coughlin, Brugamyer, Mullally, Quintana, Still, Thompson, Morrison, Twicken,
  Desert, Carter, Crepp, H{\'e}brard, Santerne, Moutou, Sobeck, Hudgins, Haas,
  Robertson, Lillo-Box, \& Barrado}]{Marcy:2014hr}
Marcy, G.~W., Isaacson, H., Howard, A.~W., {et~al.} 2014, ApJS, 210, 20

\bibitem[{Marois {et~al.}(2006)Marois, Lafreni{\`e}re, Doyon, Macintosh, \&
  Nadeau}]{Marois:2006df}
Marois, C., Lafreni{\`e}re, D., Doyon, R., Macintosh, B., \& Nadeau, D. 2006,
  ApJ, 641, 556

\bibitem[{Millar-Blanchaer {et~al.}(2015)Millar-Blanchaer, Graham, Pueyo,
  Kalas, Dawson, Wang, Perrin, Moon, Macintosh, Ammons, Barman, Cardwell, Chen,
  Chiang, Chilcote, Cotten, De~Rosa, Draper, Dunn, Duch{\^e}ne, Esposito,
  Fitzgerald, Follette, Goodsell, Greenbaum, Hartung, Hibon, Hinkley, Ingraham,
  Jensen-Clem, Konopacky, Larkin, Long, Maire, Marchis, Marley, Marois,
  Morzinski, Nielsen, Palmer, Oppenheimer, Poyneer, Rajan, Rantakyr{\"o},
  Ruffio, Sadakuni, Saddlemyer, Schneider, Sivaramakrishnan, Soummer, Thomas,
  Vasisht, Vega, Wallace, Ward-Duong, Wiktorowicz, \&
  Wolff}]{MillarBlanchaer:2015ha}
Millar-Blanchaer, M.~A., Graham, J.~R., Pueyo, L., {et~al.} 2015, ApJ, 811, 18

\bibitem[{Montet {et~al.}(2015)Montet, Bowler, Shkolnik, Deck, Wang, Horch,
  Liu, Hillenbrand, Kraus, \& Charbonneau}]{Montet:2015vr}
Montet, B.~T., Bowler, B.~P., Shkolnik, E.~L., {et~al.} 2015, eprint
  arXiv:1508.05945, 1508.05945

\bibitem[{Moutou {et~al.}(2015)Moutou, Lo~Curto, Mayor, Bouchy, Benz, Lovis,
  Naef, Pepe, Queloz, Santos, Segransan, Sousa, \& Udry}]{Moutou:2015fd}
Moutou, C., Lo~Curto, G., Mayor, M., {et~al.} 2015, A{\&}A, 576, 48

\bibitem[{Nesvorn{\'{y}} {et~al.}(2012)Nesvorn{\'{y}}, Kipping, Buchhave,
  Bakos, Hartman, \& Schmitt}]{Nesvorny:2012gs}
Nesvorn{\'{y}}, D., Kipping, D.~M., Buchhave, L.~A., {et~al.} 2012, Science,
  336, 1133

\bibitem[{Nielsen {et~al.}(2008)Nielsen, Close, Biller, Masciadri, \&
  Lenzen}]{Nielsen:2008kk}
Nielsen, E.~L., Close, L.~M., Biller, B.~A., Masciadri, E., \& Lenzen, R. 2008,
  ApJ, 674, 466

\bibitem[{Nielsen {et~al.}(2012)Nielsen, Liu, Wahhaj, Biller, Hayward, Boss,
  Bowler, Kraus, Shkolnik, Tecza, Chun, Clarke, Close, Ftaclas, Hartung, Males,
  Reid, Skemer, \& Alencar}]{Nielsen:2012jk}
Nielsen, E.~L., Liu, M.~C., Wahhaj, Z., {et~al.} 2012, ApJ, 750, 53

\bibitem[{Nielsen {et~al.}(2013)Nielsen, Liu, Wahhaj, Biller, Hayward, Close,
  Males, Skemer, Chun, Ftaclas, Alencar, Artymowicz, Boss, Clarke, de~Gouveia
  Dal~Pino, Gregorio-Hetem, Hartung, Ida, Kuchner, \& Lin}]{Nielsen:2013jy}
---. 2013, ApJ, 776, 4

\bibitem[{Nielsen {et~al.}(2014)Nielsen, Liu, Wahhaj, Biller, Hayward, Males,
  Close, Morzinski, Skemer, Kuchner, Rodigas, Hinz, Chun, Ftaclas, \&
  Toomey}]{Nielsen:2014js}
---. 2014, ApJ, 794, 158

\bibitem[{Patel {et~al.}(2014)Patel, Metchev, \& Heinze}]{Patel:2014dp}
Patel, R.~I., Metchev, S.~A., \& Heinze, A. 2014, ApJS, 212, 10

\bibitem[{Pearce {et~al.}(2015)Pearce, Wyatt, \& Kennedy}]{Pearce:2015je}
Pearce, T.~D., Wyatt, M.~C., \& Kennedy, G.~M. 2015, MNRAS, 448, 3679

\bibitem[{Perrin {et~al.}(2014)Perrin, Maire, Ingraham, Savransky,
  Millar-Blanchaer, Wolff, Ruffio, Wang, Draper, Sadakuni, Marois, Rajan,
  Fitzgerald, Macintosh, Graham, Doyon, Larkin, Chilcote, Goodsell, Palmer,
  Labrie, Beaulieu, De~Rosa, Greenbaum, Hartung, Hibon, Konopacky,
  Lafreni{\`e}re, Lavigne, Marchis, Patience, Pueyo, Rantakyr{\"o}, Soummer,
  Sivaramakrishnan, Thomas, Ward-Duong, \& Wiktorowicz}]{Perrin:2014jh}
Perrin, M.~D., Maire, J., Ingraham, P., {et~al.} 2014,
  Proc.{\textasciitilde}SPIE, 9147, 91473J

\bibitem[{Perryman {et~al.}(2014)Perryman, Hartman, Bakos, \&
  Lindegren}]{Perryman:2014jr}
Perryman, M., Hartman, J., Bakos, G.~{\'A}., \& Lindegren, L. 2014, ApJ, 797,
  14

\bibitem[{Pueyo {et~al.}(2015)Pueyo, Soummer, Hoffmann, Oppenheimer, Graham,
  Zimmerman, Zhai, Wallace, Vescelus, Veicht, Vasisht, Truong,
  Sivaramakrishnan, Shao, Roberts, Roberts, Rice, Parry, Nilsson, Lockhart,
  Ligon, King, Hinkley, Hillenbrand, Hale, Dekany, Crepp, Cady, Burruss,
  Brenner, Beichman, \& Baranec}]{Pueyo:2015cx}
Pueyo, L., Soummer, R., Hoffmann, J., {et~al.} 2015, ApJ, 803, 31

\bibitem[{Reyl{\'e} {et~al.}(2010)Reyl{\'e}, Delorme, Willott, Albert,
  Delfosse, Forveille, Artigau, Malo, Hill, \& Doyon}]{Reyle:2010gq}
Reyl{\'e}, C., Delorme, P., Willott, C.~J., {et~al.} 2010, A{\&}A, 522, 112

\bibitem[{Riviere-Marichalar {et~al.}(2014)Riviere-Marichalar, Barrado,
  Montesinos, Duch{\^e}ne, Bouy, Pinte, Menard, Donaldson, Eiroa, Krivov, Kamp,
  Mendigut{\'\i}a, Dent, \& Lillo-Box}]{RiviereMarichalar:2014jm}
Riviere-Marichalar, P., Barrado, D., Montesinos, B., {et~al.} 2014, A{\&}A,
  565, 68

\bibitem[{Simon \& Schaefer(2011)}]{Simon:2011ix}
Simon, M., \& Schaefer, G.~H. 2011, ApJ, 743, 158

\bibitem[{Soummer {et~al.}(2012)Soummer, Pueyo, \& Larkin}]{Soummer:2012ig}
Soummer, R., Pueyo, L., \& Larkin, J. 2012, ApJL, 755, L28

\bibitem[{van Leeuwen(2007)}]{vanLeeuwen:2007dc}
van Leeuwen, F. 2007, A{\&}A, 474, 653

\bibitem[{Wang {et~al.}(2015)Wang, Ruffio, De~Rosa, Aguilar, Wolff, \&
  Pueyo}]{Wang:2015th}
Wang, J.~J., Ruffio, J.-B., De~Rosa, R.~J., {et~al.} 2015, Astrophysics Source
  Code Library, -1, 06001

\bibitem[{Wizinowich {et~al.}(2000)Wizinowich, Acton, Shelton, Stomski,
  Gathright, Ho, Lupton, Tsubota, Lai, Max, Brase, An, Avicola, Olivier, Gavel,
  Macintosh, Ghez, \& Larkin}]{Wizinowich:2000hl}
Wizinowich, P., Acton, D.~S., Shelton, C., {et~al.} 2000, PASP, 112, 315

\bibitem[{Wolff {et~al.}(2014)Wolff, Perrin, Maire, Ingraham, Rantakyr{\"o}, \&
  Hibon}]{Wolff:2014cn}
Wolff, S.~G., Perrin, M.~D., Maire, J., {et~al.} 2014,
  Proc.{\textasciitilde}SPIE, 9147, 91477H

\bibitem[{Wu \& Murray(2003)}]{Wu:2003kl}
Wu, Y., \& Murray, N. 2003, ApJ, 589, 605

\bibitem[{Yelda {et~al.}(2010)Yelda, Lu, Ghez, Clarkson, Anderson, Do, \&
  Matthews}]{Yelda:2010ig}
Yelda, S., Lu, J.~R., Ghez, A.~M., {et~al.} 2010, ApJ, 725, 331

\bibitem[{Zuckerman {et~al.}(2001)Zuckerman, Song, Bessell, \&
  Webb}]{Zuckerman:2001go}
Zuckerman, B., Song, I., Bessell, M.~S., \& Webb, R.~A. 2001, ApJL, 562, L87

\end{thebibliography}
\end{document}